\documentclass[superscriptaddress,aps,twocolumn,amsmath,amssymb,10pt,a4paper,nofootinbib,prl,showkeys
]{revtex4-2}
\usepackage[utf8]{inputenc}
\usepackage{braket}
\usepackage{xcolor}
\usepackage{hyperref}
\usepackage{amsmath}
\usepackage{amssymb}
\usepackage{graphicx}
\usepackage{mathtools}
\usepackage{comment}
 \usepackage[export]{adjustbox}
\newcommand{\rd}{\mathrm{d}}

\renewcommand{\eqref}[1]{eq.~(\ref{#1})}

\newcommand{\Jcal}{\mathcal{J}}

\newcommand{\beq}{\begin{equation}}
\newcommand{\eeq}{\end{equation}}
\def\beqa#1\eeqa{\begin{align}#1\end{align}}

\begin{document}

\title{
Stabilization of macroscopic dynamics by fine-grained disorder in many-species ecosystems
}
\author{Juan Giral Mart\'inez}
\affiliation{Institut de Biologie de l'École Normale Supérieure, Département de Biologie, École Normale Supérieure, PSL Research University, Paris, France}
\author{Silvia De Monte}
\affiliation{Institut de Biologie de l'École Normale Supérieure, Département de Biologie, École Normale Supérieure, PSL Research University, Paris, France}
\affiliation{Max Planck Institute of Evolutionary Biology, Plön, Germany.}
\author{Matthieu Barbier}
\affiliation{CIRAD, UMR PHIM, 34090 Montpellier, France.}
\affiliation{PHIM Plant Health Institute, University of Montpellier,
CIRAD, INRAE, Institut Agro, IRD, 34090 Montpellier, France.}
\affiliation{Institut Natura e Teoria en Pirenèus, Surba, France.}

\begin{abstract}
A central feature of complex systems is the relevance and entanglement of different levels of description. For instance, the dynamics of ecosystems can be alternatively described in terms of large ecological processes and classes of organisms, or of individual species and their relations. Low-dimensional heuristic 'macroscopic' models that are widely used to capture ecological relationships -- and commonly evidence out-of equilibrium regimes -- implicitly assume that species-level  'microscopic' heterogeneity can be neglected. 
Here, we address the stability of such macroscopic descriptions to the addition of disordered microscopic interactions. We find that increased heterogeneity can stabilize collective as well as species fluctuations -- contrary to the well-known destabilizing effect of disorder on fixed points. We analytically find the conditions for the existence of heterogeneity-driven equilibria, and relate their stability to a mismatch in microscopic time scales. This may shed light onto the empirical observation that many-species ecosystems often appear stable at aggregated levels despite highly diverse interactions and large fluctuations at the species level.
\end{abstract}

\keywords{}

\maketitle

Complex systems encompass different levels of description, but interest is often focused on system-level properties that integrate microscopic details~\cite{tikhonov2021}. For instance, in classical ecological models the dynamics of natural communities is heuristically described in terms of a few functionally distinct groups~\cite{Goldford2018,Louca2018,laureto} (e.g. phytoplankton and zooplankton, producers and decomposers, etc), whose numerous composing species are \textit{de facto} held equivalent~\cite{chapin2002principles}. In such low-dimensional population models, system-level dynamics is posited heuristically, rather than derived by averaging microscopic degrees of freedom. Therefore the macroscopic equations are not guaranteed to be robust to uncharacterised microscopic heterogeneity~\cite{tikhonov2022}. Here, we address the consequences of adding such heterogeneity to systems with a defined out-of equilibrium macroscopic dynamics.

In order to illustrate this general problem, we consider at first an ecological \textit{macroscopic model} where the total abundance $N_\alpha, \alpha=1 \dots G$ of $G$ groups of species obeys the generalized Lotka-Volterra equations (gLV)~\cite{Roy2019,Barbier2021}:

\begin{equation}
    \frac{\rd N_\alpha}{\rd t} = r_\alpha N_\alpha \left(1 - N_\alpha - \sum_{\beta=1}^G \mu_{\alpha\beta} N_\beta\right) + \theta,
    \label{eq:glv_lowdim}
\end{equation}
where $r_\alpha$ and $\mu_{\alpha\beta}$ quantify the intrinsic growth rate of single groups and the interaction between pairs of groups, respectively, and $\theta\ll 1$ models the migration from outside the community. Notice that such groups are considered here functionally homogeneous, hence the abundances of single composing species are not important.

In real systems, however, species within groups are not identical, and are often extremely numerous~\cite{Gore2022}. One way to account for such realistic features is to divide each group into a large number $S_g$ of species (for a total of $S=G S_g$ species) that differ in their interaction parameters by a random, additive term. 
In such \textit{microscopic model}, the abundance $x_i$ of a species belonging to group $\alpha$, and of intrinsic growth rate $r_i$ obeys:
\begin{equation}
    \frac{\rd x_i}{\rd t} = r_i x_i \left(1 - x_i - \sum_{j=1}^{S} A_{ij} x_j\right) + \theta.
    \label{eq:glv_highdim}
\end{equation}
The effect of species $j$ in group $\beta$ on the growth of species $i$ is $A_{ij} = {\mu_{ij}}/{S_g} + {\sigma}\, \xi_{ij}/{\sqrt{S}}$, where $\mu_{ij} = \mu_{\alpha\beta}$ and $\xi_{ij}$ are independent random variables with zero mean and unit variance. We discuss below the properties of such interaction matrix, which combines a low-dimensional and a high-dimensional (so-called \textit{disordered}) component.
The scaling with $S_g$ ensures the existence of a well-defined thermodynamic limit when $S_g\rightarrow\infty$. The strength of microscopically disordered relative is controlled by the heterogeneity parameter $\sigma$. The well-studied random Lotka-Volterra model is recovered when $G=1$ and thus $\mu_{\alpha\beta} = \mu$~\cite{akjouj2024complex,Bunin2017}.\\

For concreteness, let's consider a simple system with macroscopic oscillations: a community composed of three groups with a rock-paper-scissors interaction matrix (RPS, see Appendix A) and $\theta>0$. We also set $r_i=r_\alpha=1$. The strength $\kappa$ of competition between pairs of groups determines whether they coexist at equilibrium with $N_\alpha = 1/(1.5+\kappa)$, $\alpha=1,2,3$ (for $\kappa \leq 3$) or they attain a limit-cycle.

In the microscopic model with $\sigma=0$, species within the same group synchronize after a transient, so that, in the asymptotic regime, group dynamics is identical to the macroscopic model: any species $i$ in group $\alpha$ follows the same trajectory as the mean abundance of the group, $m_\alpha(t) = S_g^{-1} \sum_{j\text{ in }\alpha} x_j(t)$. 
Such identity between the two levels of description holds generally for \eqref{eq:glv_highdim} with homogeneous groups, and is due to inter-specific interactions within groups
, which scale as $S_g^{-1}$, being weak compared to intra-specific interactions $1+A_{ii}$ and between-group interactions.

\begin{figure}
    \centering
    \includegraphics[width=\linewidth]{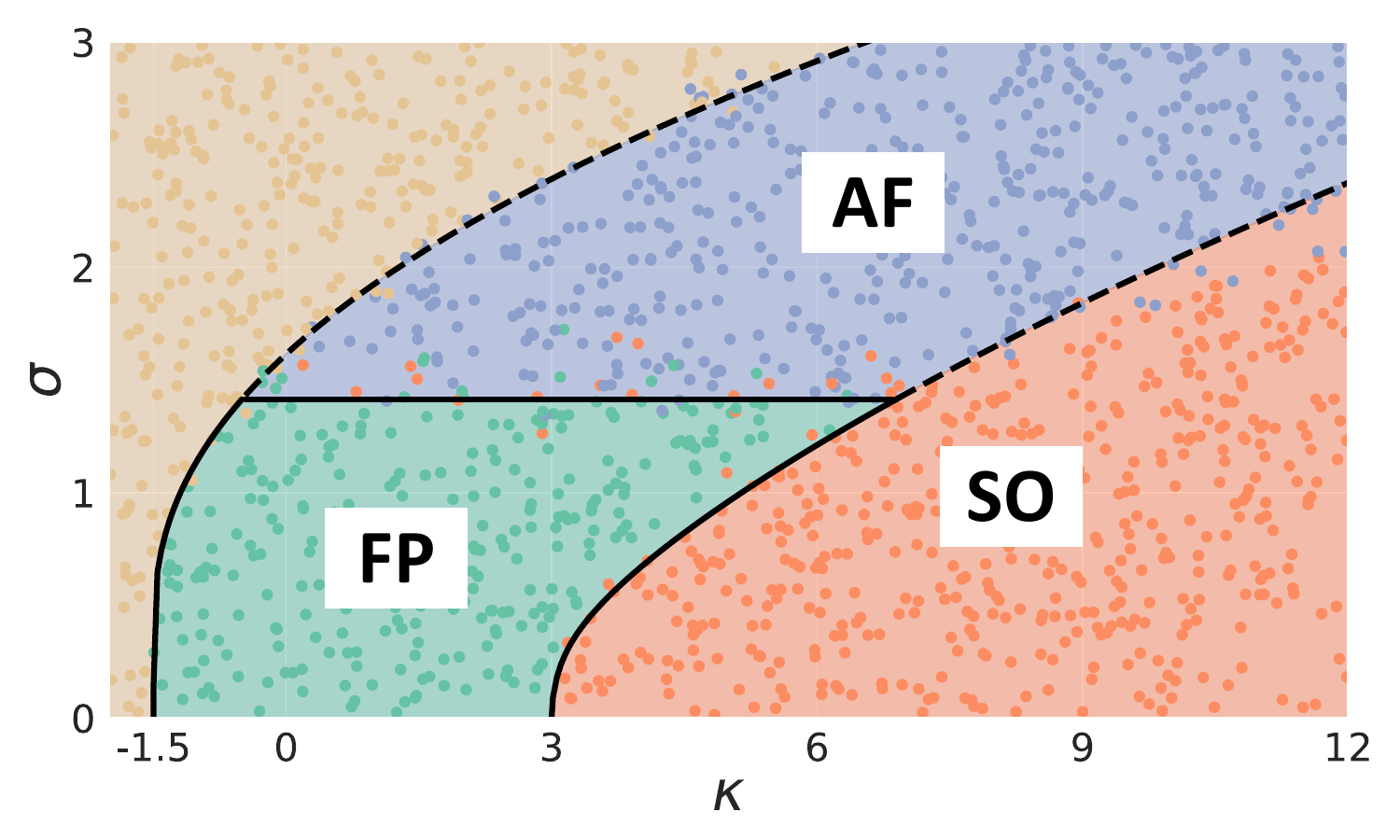}
    \caption{Phase diagram of the microscopic RPS, displaying the three meaningful phases described in the text (SO in orange, FP in green and AF in blue) as well as the physically unrealistic Unbounded Growth phase (yellow)~\cite{Bunin2017}. The SO/FP transition line was obtained using \eqref{eq: woodbury}, while the other two thick lines have already been studied in~\cite{Galla2018}. While our analysis does not apply to the SO/AF transition, a calculation assuming the fixed point solution in the blue phase yields a reasonable approximation (dashed line).}
    \label{fig:rps_phase}
\end{figure}

\begin{figure}
    \centering
    \includegraphics[width=\linewidth]{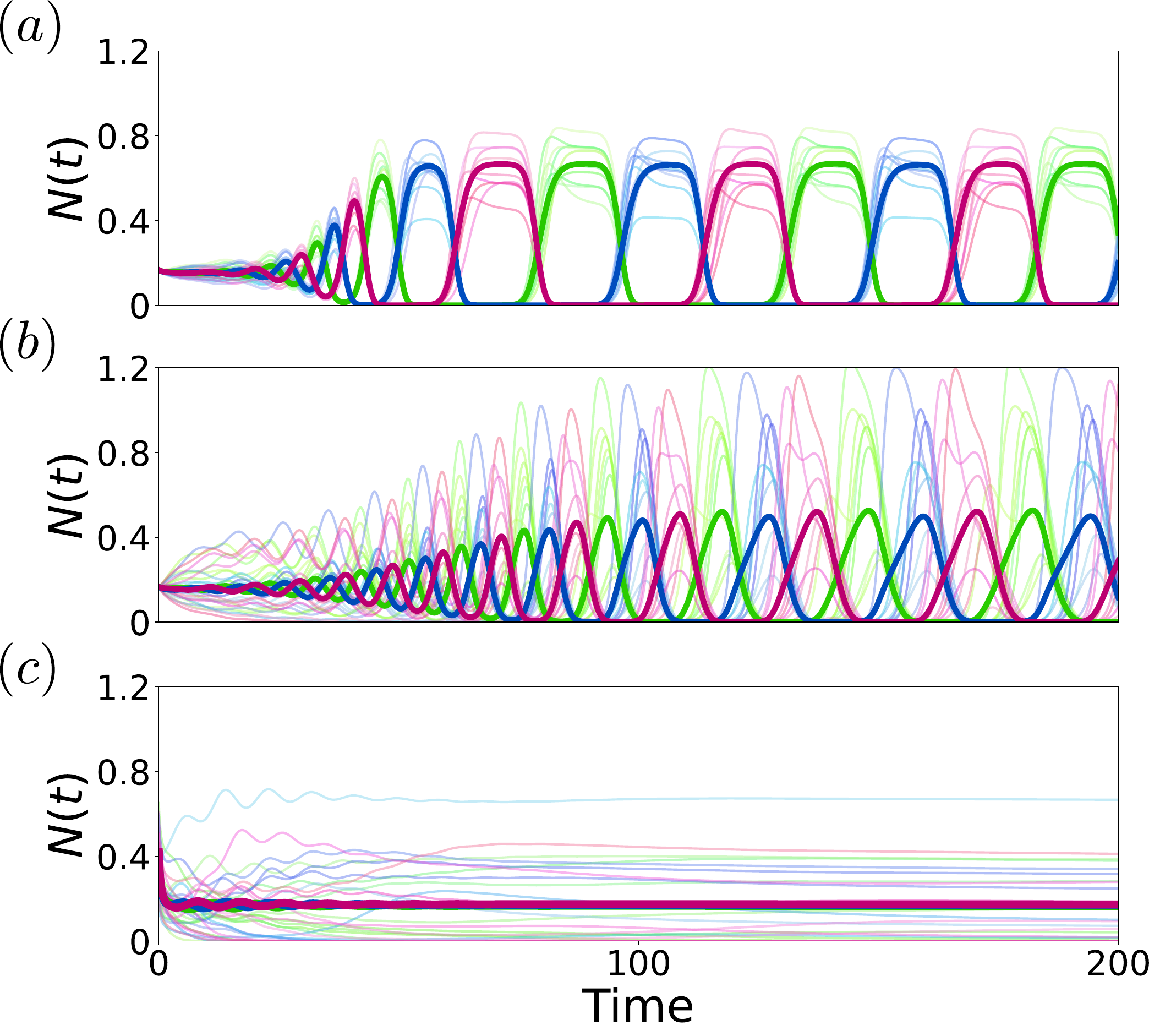}
    \caption{Trajectories for the microscopic RPS system in the SO phase. Thin lines are a subset of species's trajectories, color-coded as the groups they belong to, whose average abundance is represented as thick lines. As heterogeneity increases from $\sigma = 0.25$ (a) to $0.75$ (b), synchronous trajectories become less coherent, the amplitude of the group oscillations decreases and the transient becomes longer. Finally, the system develops a stable fixed point, seen here at $\sigma=1.1$ (c). In this simulation, $\kappa=5$ and $S_g=300$.}
    \label{fig:rps_traj}
\end{figure}

When species within groups are not identical, the dynamics possesses three meaningful phases, represented in Fig. \ref{fig:rps_phase} as a function of competition strength $\kappa$ and heterogeneity $\sigma$. In the synchronous oscillations (SO) phase, species oscillate coherently with the mean abundance $m_\alpha(t)$ of their group, and thus have the same qualitative dynamics as $N_\alpha(t)$ in the macroscopic RPS (Fig. \ref{fig:rps_traj}a). Upon increasing heterogeneity, fluctuations of species within a group become increasingly dephased, and the amplitude of group-level oscillations decreases (Fig. \ref{fig:rps_traj}b). In the fixed point phase (FP), both species and group abundances converge to a unique and stable fixed point, where all species coexist. In the asynchronous fluctuations phase (AF), species fluctuate incoherently, and apparently chaotically, so that their fluctuations cancel each other at the group level, while $m_\alpha(t)$ displays finite-size fluctuations~\cite{Roy2019}.

The transition FP/AF is akin to the disorder-induced instability of random Lotka-Volterra equations~\cite{Opper1989,Bunin2017,arnoulx2024many,arnoulx2023aging}. Since the work of May~\cite{May1972}, this transition has been a basis for the paradigm that heterogeneity has a destabilizing effect on microscopic dynamics. Of the two transitions -- SO/FP and SO/AF -- that result from the interplay between macroscopic structure and microscopic disorder, we focus on the former, where the existence of a fixed point allows us to perform a linear stability analysis.

A necessary condition for stability is that the fixed point is uninvadable. This means that extinct species cannot reinvade through migration (their growth rates $\dot{x_i}/x_i$ in \eqref{eq:glv_highdim} are negative). Using Dynamical Mean Field Theory (DMFT)~\cite{Roy2019,Galla2018} to replace \eqref{eq:glv_highdim} by effective single-species stochastic equations (see Appendix B), we find that at most one such fixed point exists. There, all groups coexist symmetrically, i.e. $m_1 = m_2 = m_3\equiv m$. When $\sigma=0$, $m$ is the equilibrium value for the macroscopic system. For $\sigma>0$, instead, the equilibrium abundances of species are distributed as
\begin{equation}
    x_i =  (1- \Tilde{\kappa} m)\;\mathrm{max}\Big(0,1 + \sqrt{\eta} \xi_i\Big),
    \label{eq:x_eq}
\end{equation}
 where $\Tilde{\kappa}=\kappa+1/2$, $\xi_i$ are independent standard Gaussian variables and $\eta = \sigma^2 q/(1-\Tilde{\kappa} m)^2$, with $q = S^{-1}\sum_i\langle (x_i)^2\rangle$, is the (rescaled) variance of abundances. The equilibrium order parameters $m,q, \eta$ can be self-consistently determined from \eqref{eq:x_eq}. In particular, one shows (see Appendix B) that $\eta$ is an increasing function of $\sigma$ only, so that higher heterogeneity results in broader abundance distributions.

\begin{figure}[h]
    \centering
    \includegraphics[width=\linewidth]{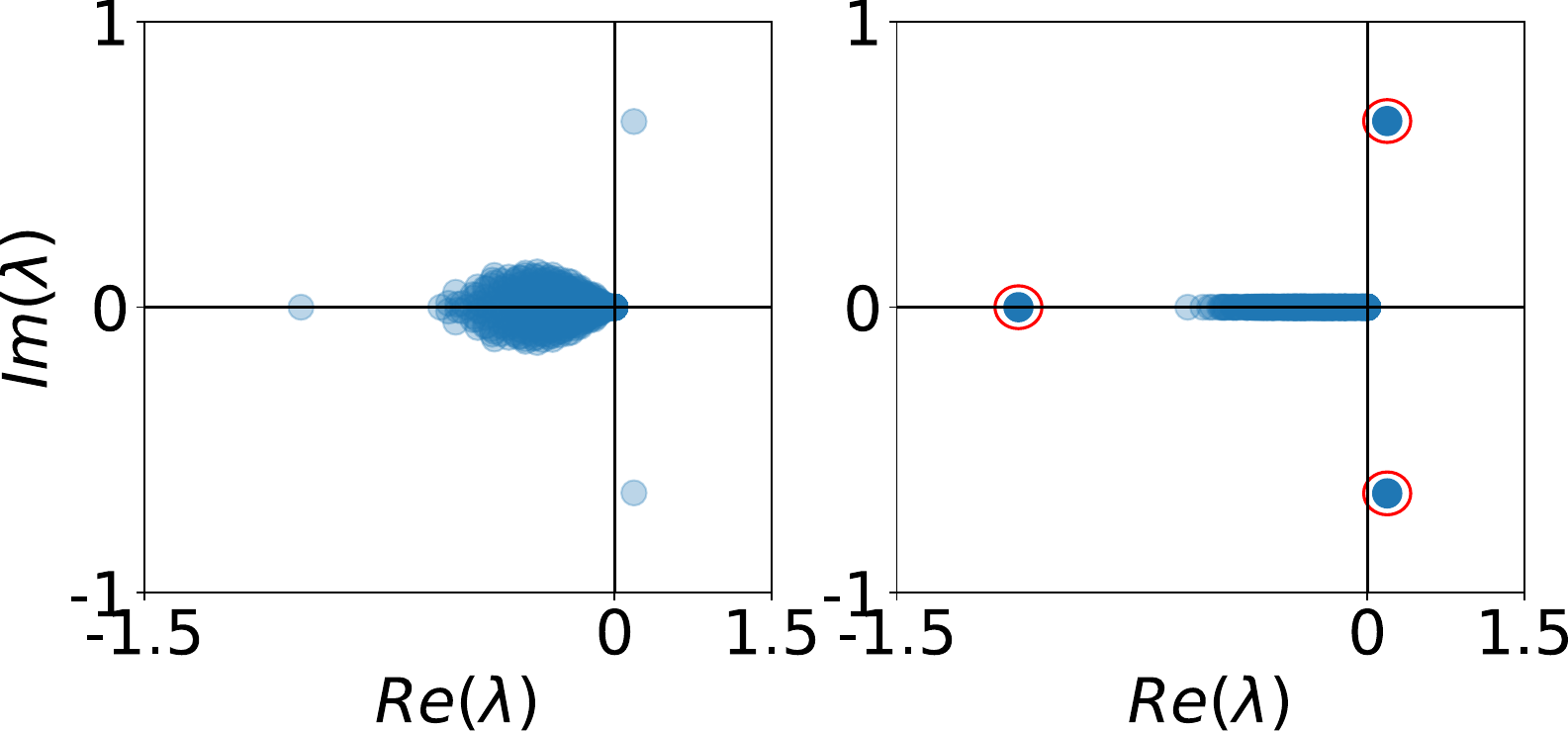}
    \caption{Spectrum of the Lotka-Volterra Jacobian $J$ (left) and the pseudo-Jacobian of \eqref{eq:Jcal} (right) for the RPS system with parameters $\kappa=5$ and $\sigma=0.75$. The red circles are predictions using Woodbury's identity \eqref{eq: woodbury}. On top of such outliers, the pseudo-Jacobian features a bulk of stable eigenvalues, corresponding to its diagonal component.}
    \label{fig:rps_spectra}
\end{figure}

\begin{figure}
    \centering
    \includegraphics[width=.9\linewidth]{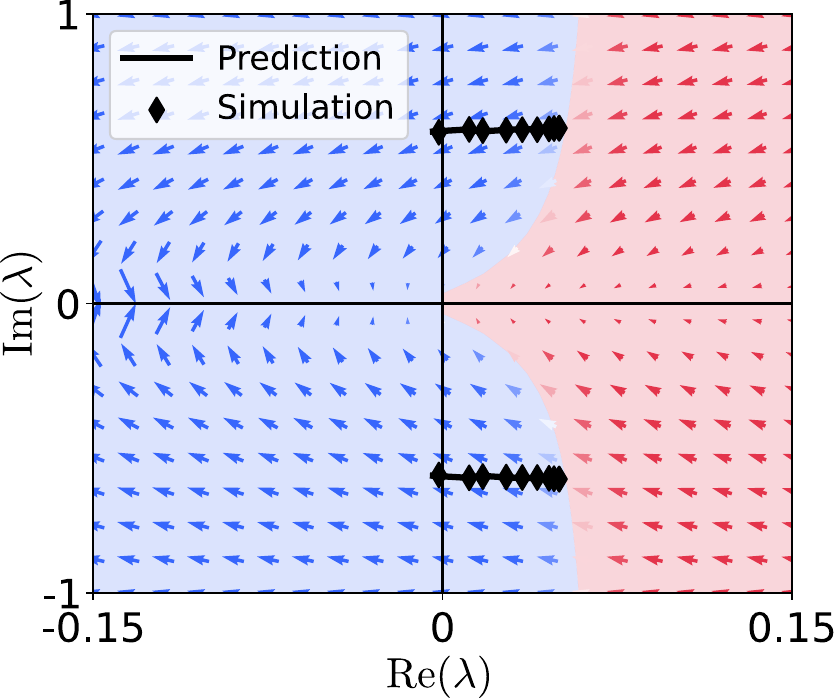}
    \caption{Flow of eigenvalues for the RPS system, as predicted by \eqref{eq:lambda_evol}. The arrows correspond to the flow at $\sigma=0$. The initial eigenvalue $\lambda_0$ (whose location depends on $\kappa$) moves left-wards as disorder increases, thus enhancing stability. The black line shows the evolution of the eigenvalues until stability is lost, for $\sigma=0.5$. The value of $\kappa=3.5$ was chosen so that the impact of extinctions in \eqref{eq: woodbury} is negligible. Dynamical simulations (diamonds) display an excellent agreement with the predictions of \eqref{eq:lambda_evol}. The red and blue regions denote eigenvalues $\lambda_0$ that are stabilized or not, respectively, when $\sigma\leq0.5$. }
    \label{fig:rps_flow}
\end{figure}

For this fixed point to be stable, it must also be robust to perturbations. This can be assessed through the eigenvalues of the Jacobian $J_{ij} = - r_i x_i \left(\delta_{ij} + A_{ij}\right)$. As is typical for large matrices, the spectrum of $J$ features a bulk and some outliers located at deterministic positions~\cite{tao2013outliers} (see Fig.\ref{fig:rps_spectra}a). 
Destabilization of the bulk entails the FP/AF transition, that has therefore the same collective character -- with many modes diverging at once -- as the May transition~\cite{bunin2016interaction}. The FP/SO transition is on the other hand associated to a bifurcation of the outliers, whose existence reflects the group structure of $\mu$. By using DMFT again, we show in Appendix C that these outliers bifurcate concomitantly to those of the simpler \textit{pseudo-Jacobian matrix} (see Fig. \ref{fig:rps_spectra}b),
\begin{equation}
    \Jcal_{ij} = - z_i\left(\delta_{ij} + \frac{\mu_{ij}}{S_g}\right),
    \label{eq:Jcal}
\end{equation}
which no longer contains the disordered part of the interactions. The variables $z_i = {r_i x_i}/(1-\Tilde{\kappa} m)$ are effective growth rates close to the fixed point, rescaled so as to be independent from the mode of the abundance distribution in \eqref{eq:x_eq}. For $\sigma=0$, $\Jcal$ has, up to the rescaling, the same outliers as the Jacobian of the macroscopic system, given by $J^M_{\alpha\beta} = -r_\alpha N_\alpha\left(\delta_{\alpha\beta} + \mu_{\alpha\beta}\right)$. Thus the stability of the microscopic and macroscopic systems is the same. As $\sigma$ is increased, each outlier $\lambda_0$ of $J^M$ is continuously modified into an outlier $\lambda(\sigma)$ of $\Jcal$ through the effect that disorder has on the $z_i$. In Appendix D, we show that in the RPS system $\lambda_0$ and $\lambda(\sigma)$ are related by
\begin{equation}
    \phi(\sigma) \left\langle \frac{1+\lambda_0}{1+z^{-1}\lambda(\sigma)}\right\rangle_{z>0} = 1,
    \label{eq: woodbury}
\end{equation}
where $\phi(\sigma)$ is the fraction of surviving species in \eqref{eq:x_eq} and the average is conditioned on $z_i>0$. When $\sigma=0$, $z_i=1$ and $\phi=1$, thus $\lambda(\sigma)=\lambda_0$. Increasing $\sigma$ in \eqref{eq: woodbury} changes $\phi$ through extinctions, resulting in a decrease of $\mathrm{Re}~\lambda(\sigma)$. Indeed, communities with fewer surviving species tend to experience weaker interactions. 
This effect can however be compensated if interactions are rescaled by the number of extant species. 
We can concentrate, instead, on the non-trivial effect of increasing abundance heterogeneity on the average in \eqref{eq: woodbury} by setting $\phi=1$. By differentiation of \eqref{eq: woodbury} with respect to $\sigma$ (see Appendix D), we obtain 
\begin{equation}
    \frac{1}{\eta'(\sigma)}\frac{\rd \lambda}{\rd \sigma} =  -\lambda \int_0^\infty \frac{ \rho_{\eta(\sigma)}(z)~\rd z}{(\lambda +z)^3} \left(\int_0^\infty \frac{\rho_{\eta(\sigma)}(z)z~\rd z}{(\lambda +z)^2}\right)^{-1}
    \label{eq:lambda_evol}
\end{equation}
where $\rho_\eta(z) = (2\pi\eta)^{-1/2}\exp\left(-(z-1)^2 / 2\eta\right)$ is the distribution of $z$. For a real and positive $\lambda$, the integrals are positive, so that the eigenvalue decreases with $\sigma$. For large heterogeneity, it converges to zero but cannot become negative, since $\mathrm{det}\left(\mathcal{J}\right) \propto \mathrm{det}(x) \mathrm{det}(\mathbb{I} + \mu/S_g)$ vanishes, in the absence of extinctions, only if $\lambda_0=0$. Complex eigenvalues, on the other hand, can cross the imaginary axis through a generic Hopf bifurcation. In Fig. \ref{fig:rps_flow}, we plot the flow associated to the r.h.s of \eqref{eq:lambda_evol} at $\sigma=0$: the real part of any complex eigenvalue will decrease with increasing $\sigma$. 
The excellent agreement of the numerical and predicted trajectories of a pair of eigenvalues of the RPS system illustrates how heterogeneity induces stabilization.

The suppression of oscillations is hence a consequence of the spread, caused by disorder, in the equilibrium abundances \eqref{eq:x_eq}. We emphasize that, having rescaled them in \eqref{eq:Jcal} and neglected extinctions in \eqref{eq: woodbury}, stabilization is only due to an increase of the variance of $x_i$. This transition reminds us of the amplitude death phenomenon in populations of synchronous oscillators, which occurs for large mismatches of intrinsic frequencies~\cite{saxena2012amplitude}, although in our system microscopic variables do not intrinsically oscillate. Indeed, as per the structure of the Jacobian \eqref{eq:Jcal}, $x_i$ controls the time scale of species $i$ around the fixed point. Increasing its heterogeneity causes a spread in the phase of species oscillations within each group. This in turn forces the amplitude of $m_\alpha(t)$ to decrease and eventually to the synchronous loss of oscillations at both the microscopic and the macroscopic scales. Ultimately, stabilization is a consequence of the multiplicative nature of ecological equations, defined in terms of per capita demographic rates -- so that abundance directly sets the time scale of species variation. Finally, we note that, due to the equivalence of $r_i$ and $x_i$ in \eqref{eq:Jcal}, the same transition would also ensue if heterogeneity was introduced in the intrinsic growth rates $r_i$.\\

The calculations above apply beyond the RPS example: for any system such that $\Vec{1}$ is an eigenvector of $\mu$, microscopic disorder is stabilizing and the bifurcation condition only depends on $\lambda_0$ and $\mu \cdot \Vec{1}$, while the other eigen-directions of $\mu$ are irrelevant. Systems with this symmetry possess a fixed point that is invariant by permutation of groups and thus depends only on the mean abundance $m = S^{-1}\Vec{x}\cdot \Vec{1}$ and the eigenvalue associated to $\Vec{1}$ ($\Tilde{\kappa}$ for the RPS). In particular, the flow of Fig. \ref{fig:rps_flow} is identical for all such fixed points. Whenever this symmetry does not hold, the equilibrium distributions differ between species. The stability of fixed points can still be assessed using \eqref{eq:Jcal}, but following the same logic as Appendix D yields a condition  more complicated than \eqref{eq: woodbury}. Although no simple criterion seems to define when the resulting flow is stabilizing, this is what we observed by numerical simulations with different matrices $\mu$ yielding out-of-equilibrium macroscopic regimes.

\begin{figure}[h]
    \centering
    \includegraphics[width=\linewidth]{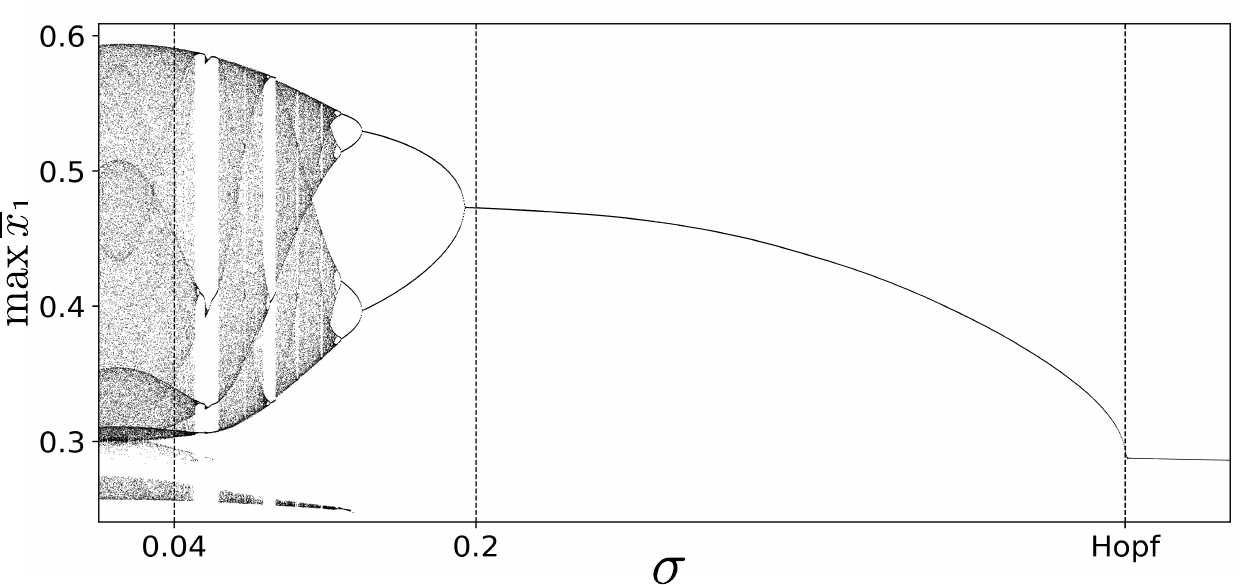}
    \caption{Poincaré map of a system with chaotic macroscopic dynamics (Appendix A) when changing heterogeneity. For small $\sigma$ (\textit{e.g.} $\sigma=0.04$), both the microscopic and macroscopic systems have chaotic dynamics. Chaos is lost through an inverse Feigenbaum cascade, it becomes cyclic (\textit{e.g.} $\sigma=0.2$) and ultimately reaches a fixed point through a Hopf bifurcation.}
    \label{fig:chaos-bif}
\end{figure}

Disorder moreover can play a stabilizing role even when the macroscopic equations do not possess a fixed point attractor, so that linear stability of equilibria should not in principle be relevant. As an example, consider a macroscopic system composed of four chaotically oscillating groups~\cite{Vano2006}, whose interaction matrix and growth rates are given in Appendix A. Increasing $\sigma$ in the microscopic model makes the oscillations undergo an inverse a Feigenbaum cascade (Fig. \ref{fig:chaos-bif}), where chaos is destroyed by a sequence of period-halving bifurcations, eventually leading to a Hopf bifurcation. The ability of microscopic heterogeneity to stabilize the community dynamics may be again connected to the spreading of time scales, even far from equilibria \cite{demonte_2002}, but this point requires further investigation. \\

In conclusion, using low-dimensional models to describe the effects of non-linearities on macroscopic  observables is common practice in ecology and other complex systems. However, this approach ignores the effect of fine-grained heterogeneity on the dynamics. In this Letter, we have embedded the macroscopic description in a more detailed microscopic one. This creates an interplay between the two levels of description that can only sustain macroscopic oscillations if the microscopic-variables remain sufficiently coherent. Increasing disorder thus leads to a series of low-codimensional bifurcations that tend to suppress macroscopic fluctuations.\\
The change of fixed point stability is the simplest example of such bifurcations. We have shown that quenched disorder results in a mismatch in the intrinsic timescales of the microscopic degrees of freedom. As their dynamics become less coherent, the amplitude of macroscopic fluctuations decreases, in a phenomenon similar to the amplitude death phenomenon observed in strongly coupled oscillators\cite{de2003coherent,kaluza2010role,saxena2012amplitude,zou2018amplitude,demonte_2002}. While we have only characterized the location of such transition for group-symmetric systems, numerical simulations show that stabilization occurs also in more general situations, and that stabilization can indeed involve transitions between out-of equilibrium regimes reminiscent of the bifurcations cascades observed for low-dimensional dynamical systems. The exploration of such transitions is an interesting future avenue of investigation. Similarly, future work should address how the SO/AF transition stems from an interference between the macroscopic dynamics and the unstable fixed points explored in~\cite{ros, arnoulx2024many}. Finally, we expect our results to be relevant to other fields where multiplicative growth is expected, such as economics~\cite{gabaix1999zipf}.

Our study raises a fundamental question: what is the scale of aggregation most relevant for addressing stability? 
Indeed, disorder has long been known to have a destabilizing effect on microscopic dynamics~\cite{May1972} through the FP/AF transition. However, this transition
is asynchronous and does not generate macroscopic fluctuations beyond finite-size effects. Thus, only synchronous transitions are relevant for macroscopic fluctuations, and we have shown that, for these, disorder can have a stabilizing effect. This is particularly relevant in ecology. When the functioning and resilience of an ecological community depends on the stability of a few macroscopic observables such as the abundance of functional groups \cite{tikhonov2021}, our model suggests that fine-grained heterogeneity could make communities more robust. On the contrary, we expect ecosystems whose function depends on a small set of species (so-called 'keystone' \cite{Doak}) to be fragile to heterogeneity-induced extinctions. Recently,  high-throughput methods for acquiring data on ecological and especially microbial communities have revealed ubiquitous fluctuations and heterogeneity at species and sub-species levels~\cite{grilli2020macroecological,martin-platero,rogers2023,Gore2022}. Yet, more macroscopic descriptions, based on large taxonomic groups or ecological functions, are markedly stable across systems and over time~\cite{Louca2016,Goldford2018,fuhrman,mutshinda}. It remains to be seen whether this macroscopic stability has other causes (e.g. physico-chemical constraints or simple aggregation effects) or truly results from species-level disorder damping ecosystem-level dynamics.

Our results demonstrate that, in order to understand the dynamics of a complex system, it is useful to discern its behaviour at different levels of description simultaneously. If this may be difficult in many practical situations, microbial communities -- where the dynamics can be assessed at the level of functional groups, but also down to the strain level~\cite{goyal2022interactions, desai} -- offer a particularly attractive opportunity for determining what kind of heterogeneity-induced transitions are most relevant for real systems.

\begin{acknowledgments}\textbf{Acknowledgments:}
We thank Guy Bunin and Jean-François Arnoldi for inspiration and discussions, and Emil Mallmin for sharing his code. J.G.M. acknowledges the support of the Frontiers in Innovation, Research and Education program. S.D.M. was supported by the French Government under the program Investissements d’Avenir (ANR-10-LABX-54 MEMOLIFE and ANR-11-IDEX- 0001-02PSL). This research was co-funded by the European Union (GA\#101059915 - BIOcean5D).
\end{acknowledgments}
\bibliography{main}
\appendix
\section{A. Parameters of the macroscopic models}
\label{appendix:macro_matrices}
The matrix $\mu$ and the growth rates for the macroscopic RPS system are
\begin{equation*}
r=
\begin{pmatrix}
    1 \\
    1\\
    1\\
    1
\end{pmatrix} , \qquad
    \mu = \begin{pmatrix}
        0.5 & \kappa & 0\\
        0 & 0.5 & \kappa\\
        \kappa & 0 & 0.5
    \end{pmatrix}.
\end{equation*}
The intra-group competition $0.5$ changes the location of the transitions, but the calculations can be done without it. The matrix $\mu$ and the growth rates for the macroscopic chaotic system are
\begin{equation*}
r=
\begin{pmatrix}
    1 \\
    0.72\\
    1.53\\
    1.27
\end{pmatrix} , \qquad
    \begin{pmatrix}
        0 & 1.09 & 1.52 & 0\\
        0 & 0 & 0.44 & 1.36\\
        2.33 & 0 & 0 & 0.47\\
        1.21 & 0.51 & 0.35 & 0.
\end{pmatrix}
\end{equation*}

In both cases, we introduce no heterogeneity in the $r_i$ when building the microscopic model, that is $r_i=r_\alpha$ for any $i$ in group $\alpha$.

\section{B. Obtaining the abundance distribution}

The evolution equation for species $i$ in group $\alpha$ can be written
\begin{equation*}
    \frac{\rd x_i}{\rd t} = r_i x_i\left(1-x_i-\sum_\beta \mu_{\alpha\beta} m_\beta(t) + \frac{\sigma}{\sqrt{S}}\sum_j \xi_{ij} x_j\right) + \theta,
\end{equation*}
where we have introduced the group abundances $m_\alpha(t) = S_g^{-1} \sum_{j \in \alpha} x_j(t)$. The well-established framework of Dynamical Mean Field Theory~\cite{Opper1989,Galla2018,Giral2024} can be used to replace the random term in this equation by an effective Gaussian noise correlated in time. Hence one obtains an effective SDE for each species,
\begin{equation}
    \frac{\rd x_i}{\rd t} = r_i x_i \left(1-x_i - \sum_\beta \mu_{\alpha\beta} m_\beta(t) + \sigma \zeta_i(t)\right) + \theta,
    \label{eq:app-dmft}
\end{equation}
where $\zeta_i$ are Gaussian processes with zero mean and correlation $\langle\zeta_i(t)\zeta_j(s)\rangle = \delta_{ij}S^{-1}\sum_i\langle x_i(t) x_i(s) \rangle$. The averages are w.r.t. the stochastic processes in \eqref{eq:app-dmft}. In particular, conditional to $m_\alpha(t)$, these are independent processes and thus the law of large numbers implies $m_\alpha(t) = S_g^{-1} \sum_{i\text{ in }\alpha} \langle x_i(t) | \{m_\beta\} \rangle$. Finally, the conditioning on the r.h.s. can be proven to be negligible due to the large number of species in each group, so that
\begin{equation*}
    m_\alpha(t) = S_g^{-1}\sum_{i\text{ in }\alpha} \langle x_i(t)\rangle
\end{equation*}
is a deterministic process. At equilibrium, we thus obtain
\begin{equation}
    x_i = \mathrm{max}\left(0,1 - \sum_\beta \mu_{\alpha\beta}m_\beta+\sigma \zeta_i\right),
    \label{eq:app_x_eq}
\end{equation}
where the species for which the abundances would be negative have been truncated to zero. In particular, this ensures that extinct species cannot invade the equilibrium. The self-consistent conditions on $q = S^{-1}\sum_i \langle x_i^2\rangle$ and $m_\alpha=S_g^{-1}\sum_{i\text{ in }\alpha} \langle x_i\rangle$ can be expressed as
\begin{equation*}
    \begin{split}
        m_\alpha &= \sigma\sqrt{q} \omega_1\left(\frac{1 - \sum_\beta \mu_{\alpha\beta} m_\beta}{\sigma\sqrt{q}}\right)\\
        1 &= \sigma^2 G^{-1}\sum_{\alpha} \omega_2\left(\frac{1 - \sum_\beta \mu_{\alpha\beta} m_\beta}{\sigma\sqrt{q}}\right)
    \end{split},
\end{equation*}
where the functions $\omega_k$ are defined as
\begin{equation}
    \omega_k (x) = (2\pi)^{-1/2}\int_{-1/\sqrt{x}}^\infty \exp(-z^2/2) (z+1/\sqrt{x})^k \rd z. 
    \label{eq:omega_def}
\end{equation}.
For matrices $\mu$ having $\Vec{1}$ as an eigenvector with eigenvalue $\kappa$, these equations have a solution where $m_\alpha=m_\beta$. This is the case of the RPS system. In that case, the group structure disappears from all the equations, and only $\kappa$ remains. Thus we obtain \eqref{eq:x_eq} in the main text. For the RPS, this turns out to be the only possibly stable equilibrium, since in any two-group combination there is always one that excludes the other and in any one-group system there is always one extinct group that can invade.

\section{C. Stability conditions}
\label{appendix:stability}
Stability can be assessed by linearizing the effective process \eqref{eq:app-dmft}. Let $\delta x_i(t) = x_i(t) - x_i$ be a small perturbation to the fixed point abundances. To first order,
\begin{equation*}
\begin{split}
    \frac{\rd}{\rd t} \delta x_i &= r_i\delta x_i \left(1-x_i - \sum_\beta \mu_{\alpha\beta} m_\beta + \sigma \zeta_i\right)\\
    &+ r_i x_i \left(- \delta x_i - \sum_\beta \mu_{\alpha\beta} \delta m_\beta + \sigma \delta \zeta_i\right).
\end{split}
\end{equation*}
The first term is zero for surviving species, and it is negative for extinct species as per \eqref{eq:app_x_eq}. Thus it always induces an exponential relaxation towards zero. We focus on the second term. Given that the $\delta \zeta_i$ are Gaussian and that the equation is linear, $\delta x_i$ are Gaussian processes, which we can characterize through their mean and correlator. Focusing on divergences at the level of the correlator leads to the May transition, for which the condition is
\begin{equation*}
    \phi \sigma^2 > 1.
\end{equation*}
This transition is located at $\sigma=\sqrt{2}$ for the RPS system, since the fixed point coincides with the fixed point of previous studies.\\

Here we we are more interested in divergences associated to the averages $\langle \delta x_i(t)\rangle$, for which we obtain the equation
\begin{equation*}
    \frac{\rd}{\rd t} \langle \Vec{\delta x}\rangle = M \langle \Vec{\delta x}\rangle,
\end{equation*}
with the matrix $M_{ij} = - r_i x_i\left(\delta_{ij} + S_g^{-1} \mu_{ij}\right)$. In the main text, we have defined the pseudo-Jacobian of RPS as $\Jcal = M / (1-\Tilde{\kappa} m)$ in order to factor the modal abundance in subsequent equations. Transitions are thus signaled by any eigenvalue of $M$ (or $\Jcal$) having a vanishing real part.

\section{D. Effect of abundance spread on eigenvalues}

To study the spectrum of $\Jcal$, we can take advantage of the low-rank of $(\mu_{ij})_{i,j\leq S}$ to reduce the problem to that of inverting a small matrix. In particular, we may write $\mu_{ij} = \sum_{\alpha\beta} \epsilon_{i\alpha} \mu_{\alpha\beta} \epsilon_{j\beta}$, where $\epsilon_{i\alpha} = 1$ if $i$ is in $\alpha$ and zero otherwise, and use Woodbury's matrix identity to obtain that the eigenvalues of $\Jcal$ satisfy
\begin{equation*}
    \mathrm{det}\left(\mu^{-1} + \epsilon^T \mathrm{diag}\left\langle \frac{z_i}{\lambda(\sigma) + z_i}\right\rangle\epsilon\right) = 0,
\end{equation*}
where the matrix in the determinant is now $G\times G$. Whenever all the $x_i$ have the same distribution, as is the case in the RPS system and for all fixed points that are symmetric by group permutation, we can use $\epsilon^T \epsilon = \mathbb{I}_n$ to obtain the condition that $-\langle z/(\lambda(\sigma) +z)\rangle$ must be an eigenvalue of $\mu^{-1}$. Introducing $\langle \dots \rangle_{z>0} = \phi^{-1} \langle \dots\rangle$ finally gives \eqref{eq: woodbury}.\\

For real eigenvalues, the concavity of the function inside the brackets in \eqref{eq: woodbury} implies that $\lambda(\sigma)/\langle z\rangle_{z>0} \leq \phi(\sigma)\left(1+\lambda_0\right) - 1$. In particular, decreasing $\phi$ decreases $\mathrm{Re}~\lambda(\sigma)$ and stabilization happens for $\phi \leq 1/(1+\mathrm{Re}~ \lambda_0)$. Setting $\phi=1$ to remove the effect of extinctions, we expect the concavity inequality to be less sharp when $\sigma$ is large, so that $\lambda(\sigma) \leq \lambda_0$ and decreasing with $\sigma$.\\

The flow of \eqref{eq:lambda_evol} can be obtained by differentiating \eqref{eq: woodbury} with $\phi\rightarrow 1$. We do so by defining a set of processes $\{y_i^\sigma\}_{\sigma\geq0}$ such that $y_i^0=1$ and $y_i^{\sigma+\rd\sigma} = y_i^\sigma + \sqrt{\eta'(\sigma)} \rd W_i^\sigma$, where $W_i$ are independent Brownian motions. Conditioned on positivity, $y^\sigma_i$ has the same law as $z_i$ for a disorder strength of $\sigma$. Plugging the evolution equation for $y_i$ in \eqref{eq: woodbury} and developping to leading order in $\delta \lambda(\sigma)$ and $\delta y_i^\sigma$ we obtain
\begin{equation*}
    \delta\lambda =  -\eta'(\sigma)\lambda \left\langle \frac{1}{(\lambda +y^\sigma)^3} \right\rangle_{y^\sigma>0} \left\langle\frac{z}{(\lambda +y^\sigma)^2}\right\rangle_{y^\sigma>0}^{-1} \delta \sigma.
\end{equation*}
Rearranging and writing the averages explicitly in terms of the distribution of $y^\sigma$ gives \eqref{eq:lambda_evol}.

\end{document}